\documentclass[12pt]{article}
\title{
\vspace{-8mm}
\rightline{\small HUB--EP--98/55}
\vspace{-2mm}
\bf String Representation of \\
Field Correlators in the \\
SU(3)-Gluodynamics}  
\author{Dmitri Antonov \thanks{Phone: 0049-30-2093 7974; Fax: 0049-30-2093 
7631; E-mail address: 
antonov@physik.hu-berlin.de} \thanks {On leave of absence from the 
Institute of Theoretical and Experimental Physics, B.Cheremushkinskaya 25, 
117 218 Moscow, Russia.} and 
Dietmar Ebert \thanks{E-mail address: 
debert@physik.hu-berlin.de}\\
{\it Institut f\"ur Physik, Humboldt-Universit\"at zu Berlin,}\\
{\it Invalidenstrasse 110, D-10115 Berlin, Germany}}
\date{}
\begin{document}
\maketitle
\vspace{1mm}
\centerline{\bf {Abstract}}
\vspace{3mm}
The string representation of the Abelian projected 
$SU(3)$-gluodynamics partition function is derived by using the 
path-integral duality transformation. On this basis, we also 
derive analogous representations for the generating 
functionals of correlators 
of gluonic field strength tensors and monopole currents, which  
are finally applied to the evaluation of the corresponding 
bilocal correlators. The large distance asymptotic behaviours 
of the latter turn out 
to be in a good agreement with existing lattice data and  
the Stochastic Model of the QCD vacuum.

\newpage
\section{Introduction}

Despite a lot of efforts towards the construction of string 
representations  
of various gauge theories possessing a confining phase (for recent 
work see e.g.~\cite{prd1, prd2} 
and Refs. therein), up to now the main 
progress in the solution of this problem has been achieved only for the 
case of Abelian theories. Those included compact QED in 3D~\cite{pol}
and 4D~\cite{diamant, dia}, and the Abelian Higgs Model (AHM) 
in the normal~\cite{fort, emil} 
and dual 
phases, extended by external electrically charged 
particles~\cite{prd2}. 

As far as gluodynamics is concerned,  
the derivation and investigation of the local string 
effective action in this theory has 
been performed in Refs.~\cite{mpla, prd1} by making use of the 
Stochastic Model of the QCD vacuum~\cite{stoch, ufn}. However, in this 
approach, the main aspect of the problem of string representation 
has not been addressed, namely the task of derivation of the complete 
string partition function as an integral over string world-sheets. 
Indeed, the Stochastic Vacuum Model (SVM) enables one to derive the string 
effective action only for the world-sheet of the minimal area. 
The reason for that is that within SVM one looses the meaning of 
the path-integral average over the gluodynamics vacuum, substituting 
this average by the phenomenological one. As a consequence, it looks 
difficult to extract singularities corresponding to QCD strings out 
of the resulting vacuum correlation functions. 

In the present Letter, we shall adopt another strategy for the construction 
of the string representation of $SU(3)$-gluodynamics. It is based on the 
so-called Abelian projection method~\cite{hooft, suzuki}. 
The essence of this method is 
a gauge fixing procedure, which reduces the original gauge group 
$SU\left(N_c\right)$ to the maximal Abelian subgroup $\left[U(1)
\right]^{N_c-1}$.  
After this partial gauge fixing, the diagonal elements of the 
matrix-valued 
vector potential transform under the remaining maximal Abelian subgroup 
as Abelian gauge fields, 
whereas the off-diagonal elements transform as charged 
matter fields. 
Such a gauge fixing diagonalizes a certain (composite) operator, 
which transforms by the adjoint representation of $SU\left(N_c\right)$.  
The singularities, which might occur if some of the eigenvalues of 
this operator coincide, then play the role of magnetic monopoles 
w.r.t. the ``photon'' fields of diagonal elements of the vector potential. 

In particular, for the $SU(2)$ case, there exists only 
one type of monopoles,  
and thus the Abelian projected 
$SU(2)$-gluodynamics is simply the dual AHM with monopoles, whereas 
in the $SU(3)$ case there emerge three types of monopoles, and the 
Abelian projection leads to a more sophisticated dual model. 
The resulting effective 
Lagrangian of this model 
constructed in Ref.~\cite{suzuki} will play a central role 
in our further investigations. This Lagrangian is also of the AHM type 
and therefore 
possesses singularities 
similar to the Abrikosov-Nielsen-Olesen strings~\cite{abrikosov} 
in dual AHM.  
These singularities could be carefully accounted for on the 
path-integral level 
by making use of the corresponding duality transformation~\cite{lee}.
(Similar calculations have been recently also performed 
in Ref.~\cite{maxim}.) 

In this Letter,   
our main goal will be the construction 
of the string representation for the correlators of gluonic field  
strength tensors and monopole currents. 
These correlators play a central role in SVM and are extremely important 
for the calculation of various QCD processes, 
which include nonperturbative 
effects. As it was mentioned above, up to now these correlators were 
considered {\it in QCD} only on the phenomenological level.
Their analytical evaluation leading to the corresponding string 
representations occurred to be possible for 
the case of the dual AHM with external electrically charged particles and  
has been performed in Ref.~\cite{prd2}. In the present Letter, we shall 
apply the Abelian projection method and the path-integral duality 
transformation to the construction of string representations for these 
correlators directly in the dual model of $SU(3)$-gluodynamics. 
Our calculations will thus partly 
parallel, but also generalize those of Ref.~\cite{prd2}. 

The outline of the Letter is as follows. In Section 2, we shall 
construct the string representation for the partition function of the 
Abelian projected $SU(3)$-gluodynamics. In Section 3, we shall 
construct analogous representations for the generating functionals 
of the gluonic field and monopole current correlators and then by making 
use of them obtain string representations for the bilocal correlators.  
Studying the large distance asymptotic behaviours of the latter we 
find them to be quite similar to those, measured in the 
lattice experiments in Ref.~\cite{DiGiac}. In particular, we find 
that the correlation length of the vacuum in SVM corresponds to the 
inverse mass of two diagonal dual gluons, which arise when the 
Abelian projection is performed. 

\section{String Representation for the Partition Function of the 
Dual Infrared SU(3)-Gluodynamics}

In the absence of quarks, the partition function of the 
infrared effective model of confinement reads~\cite{suzuki} 
\footnote{Inclusion of quarks is straightforward and can be done 
along the lines of Ref.~\cite{prd2}. However, similarly to AHM, 
those are unimportant for the problem of string representation of 
field correlators under study.}

$$
{\cal Z}=\int {\cal D}\vec B_\mu{\cal D}\chi_a\delta\left(
\sum\limits_{a=1}^3\theta_a\right)\exp\Biggl\{-\int d^4x
\Biggl[\frac14\vec F_{\mu\nu}^2+\sum\limits_{a=1}^3\Biggl[
\frac12\left|\left(
\partial_\mu-ig\vec\varepsilon_a\vec B_\mu\right)\chi_a\right|^2
+\lambda\left(\left|\chi_a\right|^2-\eta^2\right)^2\Biggr]\Biggr]\Biggr\},
\eqno (1)
$$
where $\vec F_{\mu\nu}=\partial_\mu \vec B_\nu-\partial_\nu 
\vec B_\mu$ denotes the field strength tensor of the Abelian 
vector potential $\vec B_\mu\equiv\left(B_\mu^3,B_\mu^8\right)$. 
These two (magnetic) fields, which are dual to the usual gluonic 
fields $A_\mu^3$ and $A_\mu^8$, acquire a mass $m=\sqrt{\frac{3}{2}}
g\eta$ due to the Higgs mechanism.   

Next, in Eq. (1), $\chi_a=\left|\chi_a\right|{\rm e}^{i\theta_a}$, 
$a=1,2,3$,   
are three complex scalar fields of monopoles possessing magnetic 
charges $g\vec \varepsilon_a$, respectively. Here, 

$$\vec \varepsilon_1=\left(1,0\right),~ \vec\varepsilon_2=\left(-\frac12,
-\frac{\sqrt{3}}{2}\right),~ \vec\varepsilon_3=\left(-\frac12, 
\frac{\sqrt{3}}{2}\right)$$
stand for the so-called root vectors, which define the lattice at 
which monopole charges $\vec {\mbox m}$ are distributed. 
Namely, $\vec {\mbox m}=g\sum\limits_{a=1}^{3}
\zeta_a\vec\varepsilon_a$, where $\zeta_a$'s  
are some integers. Notice, that the partition 
function (1) has been derived under the assumption that the 
dominant contribution to it is brought about by the monopoles 
with the smallest magnetic charge, $\zeta_a=\pm 1$ (see e.g. discussion 
in Ref.~\cite{pol1}). The meaning of the root vectors can be better 
understood if we mention that     
after singling out of the maximal Abelian subgroup   
$U(1)\times U(1)$ by the 
redefinition of the $SU(3)$-generators $T_i\equiv\frac{\lambda_i}{2}$, 
$i=1,...,8$, 
in the following way 

$$\vec H\equiv\left(H_1,H_2\right)=\left(T_3,T_8\right),~ 
E_{\pm 1}=\frac{1}{\sqrt{2}}\left(T_1\pm iT_2\right),~
E_{\pm 2}=\frac{1}{\sqrt{2}}\left(T_4\mp iT_5\right),~
E_{\pm 3}=\frac{1}{\sqrt{2}}\left(T_6\pm iT_7\right),$$
they obey the following commutation relations 

$$\left[\vec H, E_a\right]=\vec\varepsilon_aE_a,~ 
\left[\vec H, E_{-a}\right]=-\vec\varepsilon_aE_{-a}.$$
Thus, the root vectors can be interpreted as 
structural constants 
in the so-obtained algebra. Expanding now the vector potential 
$A_\mu\equiv A_\mu^iT_i$ over the new set of generators, we see that 
these vectors define the $U(1)\times U(1)$ 
charges of off-diagonal gluons.

As it has been demonstrated in Ref.~\cite{suzuki}, due to the fact
that the trajectories of the monopoles of all three kinds are not 
independent, there should exist a constraint $\sum\limits_{a=1}^{3}
\theta_a=0$ 
relating 
the monopole fields to each other. 
We have imposed this constraint by the introduction of a  
corresponding $\delta$-function into the R.H.S. of Eq. (1).

In what follows, we shall for simplicity 
consider the model (1) in the so-called 
London limit, i.e. at $\lambda\to\infty$. In this limit, which 
corresponds to infinitely heavy monopole fields, the 
radial parts of the latter can be integrated out, and 
we are left with the following partition function

$$
{\cal Z}=\int {\cal D}\vec B_\mu{\cal D}\theta_a^{\rm sing}
{\cal D}\theta_a^{\rm reg}{\cal D}k\delta\left(\sum\limits_{a=1}^{3}
\theta_a^{\rm sing}\right)\exp\Biggl\{\int d^4x\Biggl[
-\frac14\vec F_{\mu\nu}^2-\frac{\eta^2}{2}\sum\limits_{a=1}^{3}
\left(\partial_\mu\theta_a-g\vec\varepsilon_a\vec B_\mu\right)^2+
ik\sum\limits_{a=1}^{3}\theta_a^{\rm reg}\Biggr]\Biggr\}. \eqno (2)
$$
Similarly to AHM~\cite{fort, emil, prd2},  
in the model (1), there exist   
string-like singularities (closed vortices) of the 
Abrikosov-Nielsen-Olesen 
type. That is why, in Eq. (2) we have decomposed the total phases 
of the monopole 
fields into a singular and regular part, $\theta_a=\theta_a^{\rm sing}+
\theta_a^{\rm reg}$, and imposed the constraint of vanishing of the 
sum of regular parts by introducing the integration over the 
Lagrange multiplier $k(x)$. Analogously to the dual AHM, in the model (2), 
$\theta_a^{\rm sing}$'s describe a given electric string configuration and 
are related to the world-sheets $\Sigma_a$'s of strings of three types 
via the equations 

$$\varepsilon_{\mu\nu\lambda\rho}\partial_\lambda\partial_\rho
\theta_a^{\rm sing}(x)=2\pi\Sigma_{\mu\nu}^a(x)\equiv
\int\limits_{\Sigma_a}^{}d\sigma_{\mu\nu}(x_a(\xi))\delta(x-x_a(\xi)).
\eqno (3)$$
Here, $x_a\equiv x_\mu^a(\xi)$ is a four-vector, which parametrizes 
the world-sheet $\Sigma_a$, and $\xi=\left(\xi^1,\xi^2\right)$ stands 
for the two-dimensional coordinate. 

The path-integral duality transformation of the partition function (2) is 
parallel to that of Ref.~\cite{prd2}. The only nontriviality brought 
about by the additional integration over the Lagrange multiplier occurs 
to be apparent due to the explicit form of the root vectors. Indeed, let 
us first cast Eq. (2) into the following form 

$${\cal Z}=   
\int {\cal D}\vec B_\mu
{\cal D}k{\rm e}^{-\frac14\int d^4x \vec F_{\mu\nu}^2}
\cdot$$

$$\cdot
\int{\cal D}\theta_a^{\rm sing}
\delta\left(\sum\limits_{a=1}^{3}
\theta_a^{\rm sing}\right)
{\cal D}\theta_a^{\rm reg}{\cal D}C_\mu^a\exp\Biggl\{\int d^4x
\Biggl[-\frac{1}{2\eta^2}\left(C_\mu^a\right)^2+iC_\mu^a\left(
\partial_\mu\theta_a-g\vec\varepsilon_a\vec B_\mu\right)
+ik\sum\limits_{a=1}^{3}\theta_a^{\rm reg}\Biggr]\Biggr\} \eqno (4)$$
and carry out the integration over the $\theta_a^{\rm reg}$'s. 
In this way, one needs to solve the equation $\partial_\mu C_\mu^a=k$, 
which should hold for an arbitrary index $a$. 
The solution to this equation reads

$$
C_\mu^a(x)=\frac12\varepsilon_{\mu\nu\lambda\rho}\partial_\nu
h_{\lambda\rho}^a(x)-\frac{1}{4\pi^2}\frac{\partial}{\partial x_\mu}
\int d^4y\frac{k(y)}{(x-y)^2},$$
where $h_{\lambda\rho}^a$ stands for the Kalb-Ramond field of the 
$a$-th type. Next, making use of the constraint $\sum\limits_{a=1}^{3}
\theta_a^{\rm sing}=0$, 
replacing then the integrals over $\theta_a^{\rm sing}$'s 
by the integrals over $x_\mu^a(\xi)$'s by virtue of Eq. (3) 
and omitting for simplicity the Jacobians~\cite{emil} emerging during such 
changes of the integration variables, we arrive 
at the following representation for the partition function

$${\cal Z}=
\int {\cal D}\vec B_\mu{\rm e}^{-\frac14\int d^4x
\vec F_{\mu\nu}^2}\cdot$$

$$\cdot\int{\cal D}k\exp\Biggl\{\frac{1}{4\pi^2}\int d^4xd^4y\Biggl[
-\frac{3}{2\eta^2}
\frac{k(x)k(y)}{(x-y)^2}+ig
\left(\frac{\partial}{\partial x_\mu}
\frac{k(y)}{(x-y)^2}\right)\sum\limits_{a=1}^{3}\vec
\varepsilon_a\vec B_\mu(x) 
\Biggr]\Biggr\}\cdot$$

$$\cdot\int{\cal D}x_\mu^a(\xi)\delta\left(\sum\limits_{a=1}^{3}
\Sigma_{\mu\nu}^a\right)\int{\cal D}h_{\mu\nu}^a
\exp\Biggl\{\int d^4x\Biggl[-\frac{1}{12\eta^2}
\left(H_{\mu\nu\lambda}^a\right)^2+i\pi h_{\mu\nu}^a\Sigma_{\mu\nu}^a-
\frac{ig}{2}\varepsilon_{\mu\nu\lambda\rho}\vec\varepsilon_a
\vec B_\mu\partial_\nu h_{\lambda\rho}^a\Biggr]\Biggr\},$$ 
where $H_{\mu\nu\lambda}^a=\partial_\mu h_{\nu\lambda}^a+
\partial_\lambda h_{\mu\nu}^a+\partial_\nu h_{\lambda\mu}^a$ stands 
for the field strength tensor of the Kalb-Ramond field $h_{\mu\nu}^a$. 
Clearly, due to the explicit form of the root vectors, 
the sum $\sum\limits_{a=1}^{3}\vec\varepsilon_a\vec B_\mu$ vanishes, and 
the integration over the Lagrange multiplier thus yields an 
inessential determinant factor. Notice also, that due to Eq. (3),  
the constraint 
$\sum\limits_{a=1}^{3}\theta_a^{\rm sing}=0$ resulted into a  
constraint for the world-sheets of strings of three types
$\sum\limits_{a=1}^{3}\Sigma_{\mu\nu}^a=0$. This means that actually 
only the world-sheets of two types are independent of each other, 
whereas the third one is unambiguously fixed by the demand that the above 
constraint holds.

Integrations over the dual gauge 
field $\vec B_\mu$ as well as over the Kalb-Ramond fields 
are now straightforward. Referring the reader for the details 
to Ref.~\cite{prd2} and taking into account that due to the closeness 
of the world-sheets in our case all the boundary terms vanish, we 
finally arrive at the desired string representation 

$${\cal Z}=\int{\cal D}x_\mu^a(\xi) 
\delta\left(\sum\limits_{a=1}^{3}
\Sigma_{\mu\nu}^a\right)\exp\left[
-\frac{g\eta^3}{4}\sqrt{\frac32}\int\limits_{\Sigma_a}^{}
d\sigma_{\mu\nu}(x_a(\xi))
\int\limits_{\Sigma_a}^{}
d\sigma_{\mu\nu}(x_a(\xi'))\frac{K_1\left(m\left|x_a(\xi)-x_a(\xi')\right|
\right)}{\left|x_a(\xi)-x_a(\xi')\right|}\right], \eqno (5)$$
where $K_1$ stands for the modified Bessel function. Finally, it is 
possible to integrate over one of the three world-sheets, for 
concreteness $x_\mu^3(\xi)$. This yields the expression for the 
partition function in terms of the integral over two independent 
string world-sheets

$$
{\cal Z}=\int{\cal D}x_\mu^1(\xi){\cal D}x_\mu^2(\xi)
\exp\left\{-\frac{g\eta^3}{2}\sqrt{\frac32}\left[
\int\limits_{\Sigma_1}^{}
d\sigma_{\mu\nu}(x_1(\xi))
\int\limits_{\Sigma_1}^{}
d\sigma_{\mu\nu}(x_1(\xi'))\frac{K_1\left(m\left|x_1(\xi)-x_1(\xi')\right|
\right)}{\left|x_1(\xi)-x_1(\xi')\right|}+\right.\right.$$

$$+\int\limits_{\Sigma_1}^{}
d\sigma_{\mu\nu}(x_1(\xi))
\int\limits_{\Sigma_2}^{}
d\sigma_{\mu\nu}(x_2(\xi'))\frac{K_1\left(m\left|x_1(\xi)-x_2(\xi')\right|
\right)}{\left|x_1(\xi)-x_2(\xi')\right|}+$$

$$\left.\left.
+\int\limits_{\Sigma_2}^{}
d\sigma_{\mu\nu}(x_2(\xi))
\int\limits_{\Sigma_2}^{}
d\sigma_{\mu\nu}(x_2(\xi'))\frac{K_1\left(m\left|x_2(\xi)-x_2(\xi')\right|
\right)}{\left|x_2(\xi)-x_2(\xi')\right|}
\right]\right\}. \eqno (6)$$
According to Eq. (6), in the language of the effective string theory, 
the partition function (2) has the form of two independent string 
world-sheets, which (self-)interact by the exchanges of 
massive dual gauge bosons. 

Notice, that the string tension of the Nambu-Goto term 
and the inverse bare coupling constant of 
the rigidity term, corresponding to each of the three terms in the 
string effective action standing in the exponent on the R.H.S. of 
Eq. (6) are similar to those of Ref.~\cite{prd2} and read 

$$\sigma=\pi\eta^2\ln\frac{2M}{\gamma m},$$
and 

$$\frac{1}{\alpha_0}=-\frac{\pi}{8g^2},$$
respectively. Here, $\gamma=1.781...$ is the Euler's constant, and 
$M=2\sqrt{2\lambda}\eta$ is the monopole mass following from Eq. (1), 
which serves as an UV momentum cutoff. Obviously, due to their 
nonanalyticity in $g$, 
both of 
these quantities are manifestly nonperturbative. The second 
important observation is that $\alpha_0<0$, which signifies 
on the stability of the string world-sheets~\cite{crump, dia}.

It is also worth noting, that according to Eq. (6), the energy density 
corresponding to the obtained effective nonlocal string Lagrangian  
increases not only 
with the distance between two points lying on the same world-sheet, but 
also with the distance between two different world-sheets. This means 
that also the ensemble of strings as a whole displays confining 
properties.

\section{String Representation of Field and Current Correlators}

This Section is the main part of the present Letter. Here, we shall 
derive string representations for field correlators of gluonic field 
strength tensors and monopole currents. Let us start with the string 
representation for the generating functional of the correlators of 
gluonic field strength tensors. In the London limit, this object reads 

$$
{\cal Z}\left[\vec S_{\alpha\beta}\right]=
\int {\cal D}\vec B_\mu{\cal D}\theta_a^{\rm sing}
{\cal D}\theta_a^{\rm reg}{\cal D}k\delta\left(\sum\limits_{a=1}^{3}
\theta_a^{\rm sing}\right)\cdot$$

$$\cdot\exp\Biggl\{\int d^4x\Biggl[
-\frac14\vec F_{\mu\nu}^2-\frac{\eta^2}{2}\sum\limits_{a=1}^{3}
\left(\partial_\mu\theta_a-g\vec\varepsilon_a\vec B_\mu\right)^2+
ik\sum\limits_{a=1}^{3}\theta_a^{\rm reg}+i\vec S_{\mu\nu}
\tilde{\vec F}_{\mu\nu}
\Biggr]\Biggr\}, \eqno (7)
$$
where $\vec S_{\mu\nu}$ stands for the source of the field strength tensor 
$\tilde{\vec F}_{\mu\nu}\equiv\frac12\varepsilon_{\mu\nu\lambda\rho}
\vec F_{\lambda\rho}$, which is nothing else but the field strength 
of the usual gluonic field $\vec A_\mu=\left(A_\mu^3, A_\mu^8\right)$. 
Performing with Eq. (7) the same path-integral duality transformation as 
the one applied in the previous Section, we arrive at the following 
string representation for this generating functional 

$$      
{\cal Z}\left[\vec S_{\alpha\beta}\right]=\exp\left(-\int d^4x 
\vec S_{\mu\nu}^2\right)
\int{\cal D}x_\mu^a(\xi) 
\delta\left(\sum\limits_{a=1}^{3}
\Sigma_{\mu\nu}^a\right)\cdot$$

$$\cdot\exp\left[-\int d^4xd^4y
\left(\pi\Sigma_{\lambda\nu}^a(x)-ig\vec\varepsilon_a
\vec S_{\lambda\nu}(x)\right)D_{\lambda\nu,\mu\rho}^{ab}(x-y)
\left(\pi\Sigma_{\mu\rho}^b(y)-ig\vec\varepsilon_b
\vec S_{\mu\rho}(y)\right)\right], \eqno (8)
$$
where $D_{\lambda\nu,\mu\rho}^{ab}(x)$ denotes the propagator of the 
Kalb-Ramond field $h_{\mu\nu}^a$, defined as (cf. Ref.~\cite{prd2})

$$D_{\lambda\nu, \mu\rho}^{ab}(x)\equiv
\delta^{ab}\left[
D_{\lambda\nu, \mu\rho}^{(1)}(x)+
D_{\lambda\nu, \mu\rho}^{(2)}(x)\right],$$
where

$$D_{\lambda\nu, \mu\rho}^{(1)}(x)=\frac{g\eta^3}{8\pi^2}
\sqrt{\frac32}\frac
{K_1}{\left|x\right|}\Biggl(\delta_{\lambda\mu}\delta_{\nu\rho}-
\delta_{\mu\nu}\delta_{\lambda\rho}\Biggr), $$

$$D_{\lambda\nu, \mu\rho}^{(2)}(x)=\frac{\eta}{4\pi^2gx^2}\sqrt{\frac23}
\left\{\Biggl[
\frac{K_1}{\left|x\right|}+\frac m2\left(K_0+K_2\right)\Biggr]
\Biggl(\delta_{\lambda\mu}\delta_{\nu\rho}-\delta_{\mu\nu}\delta_{\lambda
\rho}\Biggr)+\right.$$

$$+\frac{1}{2\left|x\right|}\Biggl[3\Biggl(\frac{m^2}{4}
+\frac{1}{x^2}\Biggr)K_1+\frac{3m}{2\left|x\right|}\left(K_0+
K_2\right)+\frac{m^2}{4}K_3\Biggr]\cdot$$

$$\left.\cdot\Biggl(\delta_{\lambda\rho}x_\mu x_\nu+\delta_{\mu\nu}
x_\lambda 
x_\rho-\delta_{\mu\lambda}x_\nu x_\rho-\delta_{\nu\rho}x_\mu x_\lambda
\Biggr)\right\},$$
with $K_n\equiv K_n(m|x|)$ standing for the modified Bessel functions.
In the last exponent on the R.H.S. of Eq. (8), the term 

$$-\pi^2\int\limits_{\Sigma_a}^{} 
d\sigma_{\lambda\nu}\left(x_a(\xi)\right)\int\limits_{\Sigma_a}^{} 
d\sigma_{\mu\rho}\left(x_a(\xi')\right)D_{\lambda\nu, 
\mu\rho}^{(2)}\left(x_a(\xi)-x_a(\xi')\right)$$
can be rewritten as a boundary one and thus vanishes due to the closeness 
of the string world-sheets. 

The string representation for the bilocal correlator of the 
field strength 
tensors can now immediately be read off from Eq. (8) by making use of the 
equality $\varepsilon_a^i\varepsilon_a^j=\frac32 \delta^{ij}$, where 
$i,j=1,2$ are the $U(1)\times U(1)$ indices referring to the 
generators $T_3, T_8$. We obtain

$$\left.\left<\tilde F_{\lambda\nu}^i(x)\tilde F_{\mu\rho}^j(y)\right>=
\frac{1}{{\cal Z}[0]}\frac{\delta^2{\cal Z}
\left[\vec S_{\alpha\beta}\right]}
{\delta S_{\lambda\nu}^i(x)\delta S_{\mu\rho}^j(y)}
\right|_{\vec S_{\alpha\beta}=0}=$$

$$=\delta^{ij}\left[
\left(\delta_{\lambda\mu}\delta_{\nu\rho}-\delta_{\lambda\rho}
\delta_{\mu\nu}\right)\delta(x-y)+
g^2D_{\lambda\nu, \mu\rho}^{aa}(x-y)\right]-$$

$$-4\pi^2g^2\varepsilon_a^i\varepsilon_b^j
\left<\int\limits_{\Sigma_c}^{}d\sigma_{\alpha\beta}
(x_c(\xi))\int\limits_{\Sigma_d}^{}d\sigma_{\gamma\zeta}
(x_d(\xi'))D_{\alpha\beta,\lambda\nu}^{ac}(x_c(\xi)-x)
D_{\gamma\zeta,\mu\rho}^{bd}
(x_d(\xi')-y)\right>_{x_a(\xi)}, \eqno (9)$$
where

$$\left<...\right>_{x_a(\xi)}\equiv\frac{\int {\cal D}x_\mu^a(\xi)
\delta\left(\sum\limits_{a=1}^{3}
\Sigma_{\mu\nu}^a\right)(...)\exp\left[
-\frac{g\eta^3}{4}\sqrt{\frac32}\int\limits_{\Sigma_a}^{}
d\sigma_{\mu\nu}(x_a(\xi))
\int\limits_{\Sigma_a}^{}
d\sigma_{\mu\nu}(x_a(\xi'))\frac{K_1\left(m\left|x_a(\xi)-x_a(\xi')\right|
\right)}{\left|x_a(\xi)-x_a(\xi')\right|}\right]}
{\int {\cal D}x_\mu^a(\xi)
\delta\left(\sum\limits_{a=1}^{3}
\Sigma_{\mu\nu}^a\right)\exp\left[
-\frac{g\eta^3}{4}\sqrt{\frac32}\int\limits_{\Sigma_a}^{}
d\sigma_{\mu\nu}(x_a(\xi))
\int\limits_{\Sigma_a}^{}
d\sigma_{\mu\nu}(x_a(\xi'))\frac{K_1\left(m\left|x_a(\xi)-x_a(\xi')\right|
\right)}{\left|x_a(\xi)-x_a(\xi')\right|}\right]} \eqno (10)$$
defines the average over the string world-sheets, 
and the term with the $\delta$-function 
on the R.H.S. of Eq. (9) corresponds to the free contribution to the 
correlator. 

At this point, let us recall that the theory (1) is a 
{\it large distance} effective theory of confinement. This means 
that only the large-distance asymptotic behaviours of the correlator 
(9) should be compatible with those of $SU(3)$-gluodynamics. 
At such distances, 
i.e. at $|x|\gg\frac{1}{m}$, the propagator of the Kalb-Ramond field 
$D_{\lambda\nu,\mu\rho}^{ab}(x)$ 
has the order of magnitude $g^2\eta^4$, and 
therefore the last term on the R.H.S. of Eq. (9) can be disregarded 
w.r.t. the second one, provided that the following inequality holds

$$g\eta^2\cdot\max\limits_a^{}\left|\Sigma_a\right|\ll 1, \eqno (11)$$
where $\left|\Sigma_a\right|$ stands for the area of the world-sheet 
$\Sigma_a$. 
From now on, we shall consider the case of small enough $\eta$ and/or 
$g$, for which inequality (11) takes place, and thus only the second 
term on the R.H.S. of Eq. (9) is sufficient.

Our aim below will be to compare the bilocal correlator (9) in the 
approximation (11) with 
the one of SVM, parametrized by two Lorentz structures as 
follows~\cite{stoch, ufn}

$$\left<\tilde F_{\lambda\nu}^i(x)\tilde F_{\mu\rho}^j(0)\right>=
\delta^{ij}\Biggl\{
\Biggl(\delta_{\lambda\mu}\delta_{\nu\rho}-\delta_{\lambda\rho}
\delta_{\nu\mu}\Biggr)D\left(x^2\right)+$$

$$+\frac12\Biggl[\partial_\lambda
\Biggl(x_\mu\delta_{\nu\rho}-x_\rho\delta_{\nu\mu}\Biggr)
+\partial_\nu\Biggl(x_\rho\delta_{\lambda\mu}-x_\mu\delta_{\lambda\rho}
\Biggr)\Biggr]D_1\left(x^2\right)\Biggr\}, \eqno (12)$$
i.e. to find the coefficient functions $D$ and $D_1$. 
Let us point out once more, that Eq. (12) is nothing else, but the 
correlator of two usual gluonic field strength tensors, 
$F_{\mu\nu}^3(A)$ and/or $F_{\mu\nu}^8(A)$.

Direct comparison 
of Eqs. (9) and (12) then yields

$$D\left(x^2\right)=\frac{m^3}{4\pi^2}\frac{K_1}{\left|x\right|}
\longrightarrow\frac{m^4}{4\sqrt{2}\pi^{\frac32}}
\frac{{\rm e}^{-m\left|x\right|}}{\left(m\left|x\right|\right)^
{\frac32}},~ 
|x|\gg\frac{1}{m}, 
\eqno (13)$$
and 

$$D_1\left(x^2\right)=\frac{m}{2\pi^2x^2}\Biggl[\frac{K_1}{\left|x\right|}
+\frac{m}{2}\Biggl(K_0+K_2\Biggr)\Biggr]
\longrightarrow\frac{m^4}{2\sqrt{2}\pi^{\frac32}}
\frac{{\rm e}^{-m\left|x\right|}}{\left(m\left|x\right|\right)^
{\frac52}},~ 
|x|\gg\frac{1}{m}. 
\eqno (14)$$
We now see that the obtained functions (13) and (14) coincide with 
those from Ref.~\cite{prd2}. Obviously, the bilocal correlator (12) is 
nonvanishing only for the gluonic field strength tensors of the same 
kind, i.e. for $i=j=1$ or $i=j=2$. It is straightforward to show, 
that this property remains to be valid if one accounts for the 
last term on the R.H.S. of Eq. (9).
 
Hence, for these diagonal correlators, 
the vacuum of the model (1) in the London limit 
exhibits a nontrivial correlation length, 
$T_g=\frac{1}{m}$. 
Notice also, that the large distance 
asymptotic behaviours (13) and (14) of the 
correlation functions $D$ and $D_1$ correspond to those 
obtained for the $SU(3)$-gluodynamics in the lattice 
experiments~\cite{DiGiac}. 
Namely, both of them decrease exponentially at the distance $T_g$, 
and secondly $D_1\ll D$ due to the preexponential power-like behaviours.

Finally, it is quite instructive to rederive the coefficient function (13) 
from the string representation for the correlator of two monopole 
currents, $\vec j_\mu=-g\eta^2\vec\varepsilon_a\left(\partial_\mu
\theta_a-g\left(\vec\varepsilon_a\vec B_\mu\right)\right)$. This can 
be done by virtue of the equation~\cite{ufn}

$$\left<j_\beta^i(x)j_\sigma^j(y)\right>=\delta^{ij}
\Biggl(\frac{\partial^2}
{\partial x_\mu\partial y_\mu}\delta_{\beta\sigma}-
\frac{\partial^2}{\partial x_\beta\partial y_\sigma}\Biggr)
D\left((x-y)^2\right), \eqno (15)$$
which can be obtained from the equations of motion. 
Besides that, it is also useful to derive the 
string representation for the 
generating functional of the monopole current correlators itself, which 
can then be applied to a derivation of the bilocal correlator. Such a 
generating functional reads

$$
\hat {\cal Z}\left[\vec J_\mu\right]=
\int {\cal D}\vec B_\mu{\cal D}\theta_a^{\rm sing}
{\cal D}\theta_a^{\rm reg}{\cal D}k\delta\left(\sum\limits_{a=1}^{3}
\theta_a^{\rm sing}\right)\cdot$$

$$\cdot\exp\Biggl\{\int d^4x\Biggl[
-\frac14\vec F_{\mu\nu}^2-\frac{\eta^2}{2}\sum\limits_{a=1}^{3}
\left(\partial_\mu\theta_a-g\vec\varepsilon_a\vec B_\mu\right)^2+
ik\sum\limits_{a=1}^{3}\theta_a^{\rm reg}+\vec J_\mu\vec j_\mu 
\Biggr]\Biggr\}.
$$
Once being applied to this object, path-integral duality transformation 
yields for it the following string representation

$$\hat {\cal Z}\left[\vec J_\mu\right]=\hat {\cal Z}[0]\exp\left[
\frac{m^2}{2}\int d^4x \vec J_\mu{\,}^2(x)\right]\cdot$$

$$\cdot\left<
\exp\Biggl\{g\varepsilon_{\lambda\nu\alpha\beta}
\int d^4x d^4y\Biggl[
-\frac{g}{8}\varepsilon_{\mu\rho\gamma\delta}\Biggl(\frac{\partial^2}
{\partial x_\alpha\partial y_\gamma}D_{\lambda\nu, \mu\rho}^{aa}(x-y)
\Biggr)\vec J_\beta (x) \vec J_\delta (y)+\right.$$

$$\left.+\pi\vec\varepsilon_a \Sigma_{\mu\rho}^b(y)\Biggl(
\frac{\partial}{\partial x_\alpha} D_{\lambda\nu, \mu\rho}^{ab}(x-y)\Biggr)
\vec J_\beta (x)\Biggr]\Biggr\}\right>_{x_a(\xi)}, \eqno (16)$$ 
where $\hat {\cal Z}[0]$ is defined by Eq. (5). Then, the string 
representation for the bilocal correlator following from Eq. (16), 
reads

$$\left<j_\beta^i(x) j_\sigma^j(y)\right>=
m^2\delta^{ij}\delta_{\beta\sigma}\delta(x-y)+
g^2
\varepsilon_{\lambda\nu\alpha\beta}\varepsilon_
{\mu\rho\gamma\sigma}\left\{-\frac14\delta^{ij}\frac{\partial^2}{\partial 
x_\alpha \partial y_\gamma}D_{\lambda\nu, \mu\rho}^{aa}(x-y)+\right.$$

$$\left.+\pi^2\varepsilon_a^i\varepsilon_b^j\left< 
\int\limits_{\Sigma_c}^{} d\sigma_{\delta\zeta}(x_c(\xi))
\int\limits_{\Sigma_d}^{} 
d\sigma_{\chi\varphi}(x_d(\xi'))
\Biggl(\frac{\partial}{\partial x_\alpha}D_{\lambda\nu, 
\delta\zeta}^{ac}\left(x-x_c(\xi)\right)\Biggr)\Biggl(\frac{\partial}
{\partial y_\gamma} D_{\mu\rho, \chi\varphi}^{bd}\left(y-x_d(\xi')\right)
\Biggr)
\right>_{x_a(\xi)}\right\}. \eqno (17)$$
By comparing of Eqs. (15) and (17), we recover, in the approximation (11), 
the coefficient function (13). This confirms the consistency of our 
calculations. 

\section{Summary}

In the present Letter,  
we have generalized the results of Ref.~\cite{prd2} to the non-Abelian 
case of $SU(3)$-gluodynamics, by making use of the Abelian projection 
method~\cite{hooft, suzuki}. In particular, we have derived the string 
representation for the partition function of the effective infrared 
dual model of confinement, proposed in Ref.~\cite{suzuki}, in the 
London limit. It 
turned out to have the 
form of two independent string world-sheets, which interact with 
each other 
and also self-interact by virtue of the exchanges of the massive 
dual gauge bosons. The string tension of the Nambu-Goto term and the 
inverse bare coupling constant of the rigidity term following from 
the obtained string effective action are similar to those of 
Ref.~\cite{prd2}. Namely, both of these quantities are nonanalytic 
in the magnetic coupling constant, which means their nonperturbative 
nature, and, secondly, the rigidity coupling constant turned out to be  
negative, which is sufficient for the stability of the string   
world-sheets~\cite{crump, dia}.

Finally, we have derived string 
representations for the generating functionals of correlators of 
gluonic field strength tensors and monopole currents. As it turns out, 
the vacuum of the model under study displays a nontrivial 
correlation length, which is seen in the  
correlators of diagonal (w.r.t. the $U(1)\times U(1)$ 
maximal Abelian subgroup) gluons {\it of the same kind}. 
This length is equal to the inverse mass of the dual gauge bosons. 
We have also established a correspondence to SVM and derived two 
coefficient functions, which parametrize the bilocal correlator in 
our model for the case of the above described nontrivial correlators. 
The infrared asymptotic behaviours of these functions turned out to 
be in agreement with the corresponding lattice data. Finally, by making 
use of the string representation for the bilocal correlator of the 
monopole currents and equations of motion, we have recovered the 
expression for one of the coefficient functions, thus confirming 
the consistency of the performed calculations. 

In conclusion, the proposed approach gives a new status to SVM for the 
case of $SU(3)$-gluodynamics, by 
virtue of the methods of the Abelian projection and path-integral 
duality transformation. It also provides us with some new insights 
to the solution 
of the long standing problem of string representation of QCD.

\section*{Acknowledgments}

The work of D.A. is supported by the Graduiertenkolleg 
{\it Elementarteilchenphysik} 
of the Humboldt University of Berlin. He is also grateful to
the theory group of the Quantum Field Theory Department 
of this University for their hospitality and M.I. Polikarpov for 
informing him about the existence of the paper~\cite{maxim} .

\end{document}